\documentclass[final,5p,times,twocolumn]{elsarticle}

\usepackage{graphicx}
\usepackage{amssymb}
\usepackage{subfigure}
\usepackage{threeparttable}

\journal{Physics Letters B}

\begin{document}

\begin{frontmatter}



\title{New method of precise measurement of positronium hyperfine splitting}


\author[tokyo]{A.~Ishida\corref{cor1}}
\ead{ishida@icepp.s.u-tokyo.ac.jp}

\author[tokyo]{G.~Akimoto}
\author[tokyo]{Y.~Sasaki}
\author[icepp]{T.~Suehara}
\author[icepp]{T.~Namba}
\author[tokyo]{S.~Asai}
\author[icepp]{T.~Kobayashi}
\author[komaba]{H.~Saito}
\author[kek]{M.~Yoshida}
\author[kek]{K.~Tanaka}
\author[kek]{and A.~Yamamoto}

\address[tokyo]{Department of Physics, Graduate School of Science, The University of Tokyo, 7-3-1 Hongo, Bunkyo-ku, Tokyo 113-0033, Japan}
\address[icepp]{International Center for Elementary Particle Physics (ICEPP), The University of Tokyo, 7-3-1 Hongo, Bunkyo-ku, Tokyo 113-0033, Japan}
\address[komaba]{Department of General Systems Studies, Graduate School of Arts and Sciences, The University of Tokyo, 3-8-1 Komaba, Meguro-ku, Tokyo 153-8902, Japan}
\address[kek]{High Energy Accelerator Research Organization (KEK), 1-1 Oho, Tsukuba, Ibaraki 305-0801, Japan}

\cortext[cor1]{Corresponding author (TEL:+81-3-3815-8384 / FAX:+81-3-3814-8806)}

\begin{abstract}
The ground state hyperfine splitting of positronium, $\Delta _{\mathrm{HFS}}$, 
is sensitive to high order corrections of QED. 
A new calculation up to $\mathrm{O}(\alpha ^3 \ln \alpha)$ has revealed a $3.9\sigma$ discrepancy 
between the QED prediction and the experimental results. 
This discrepancy might either be due to systematic problems in the previous experiments 
or to contributions beyond the Standard Model. 
We propose an experiment to measure $\Delta _{\mathrm{HFS}}$ employing new methods 
designed to remedy the systematic errors which may have affected the previous experiments. 
Our experiment will provide an independent check of the discrepancy. 
The prototype run has been finished and a result of 
$\Delta _{\mathrm{HFS}} = 203.380\,4 \pm 0.008\,4\,\mathrm{GHz}\,(41\,\mathrm{ppm})$ has been obtained. 
A measurement with a precision of O(ppm) is expected within a few years.
\end{abstract}


\end{frontmatter}

% \linenumbers

\section{Introduction}
\label{sec:introduction}
Positronium (Ps), a bound state of an electron and a positron, is a purely leptonic system 
which allows for very sensitive tests of Quantum ElectroDynamics (QED). 
The precise measurement of the hyperfine splitting between orthopositronium (o-Ps, 1$^3S_1$) and parapositronium (p-Ps, 1$^1S_0$) (Ps-HFS) provides a good test of bound state QED. 
Ps-HFS is expected to be relatively large (for example compared to hydrogen HFS) due to a relatively large 
spin-spin interaction, and also due to the contribution from vacuum oscillation 
(o-Ps $ \rightarrow \gamma ^{\ast} \rightarrow$ o-Ps). 
The contribution from vacuum oscillation is sensitive to new physics beyond the Standard Model. 

\begin{figure}
\begin{center}
\includegraphics[width=0.4\textwidth]{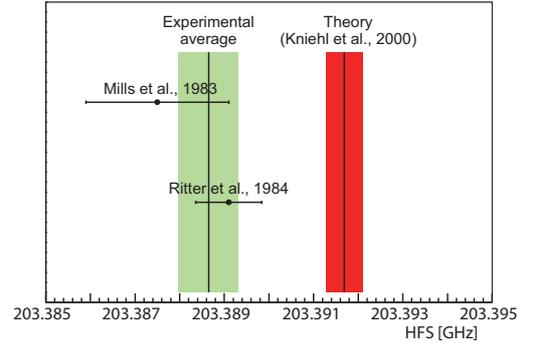}
\caption{\label{fig:history}Measured and theoretical values of Ps-HFS.}
\end{center}
\end{figure}

Figure \ref{fig:history} shows the measured and theoretical values of Ps-HFS. 
The combined value from the results of the previous 2 experiments is 
$\Delta _{\mathrm{HFS}} ^{\mathrm{exp}} = 203.388\,65(67)\,\mathrm{GHz} \,(3.3\,\mathrm{ppm})$
~\cite{HUGHES-V,MILLS-II}. 
Recent developments in NonRelativistic QED (NRQED) have added 
$\mathrm{O}(\alpha ^3 \ln \alpha)$ corrections to the theoretical prediction which now stands at 
$\Delta_{\mathrm{HFS}} ^{\mathrm{th}} = 203.391\,69(41)\,\mathrm{GHz}\,(2.0\,\mathrm{ppm})$~\cite{HFS-ORDER3}. 
The discrepancy of 3.04(79)\,MHz (15\,ppm, 3.9$\sigma$) between $\Delta _{\mathrm{HFS}} ^{\mathrm{exp}}$ 
and $\Delta_{\mathrm{HFS}} ^{\mathrm{th}}$ might either be due to the common systematic uncertainties in the previous experiments or to 
new physics beyond the Standard Model. 

There are two possible common systematic uncertainties in the previous experiments. 
One is the unthermalized o-Ps contribution which results in an underestimation of the material effect. 
This effect has already been shown to be significant~\cite{KATAOKA, KATAOKA-D, ASAI} in the 
o-Ps lifetime puzzle. 
The other is the uncertainty in the magnetic field uniformity which was cited as 
the most significant systematic error by previous experimenters. 

\section{Theory of Experiment}
\label{sec:theoryofexperiment}
\subsection{Measurement using Zeeman effect}
\label{sec:measurementusingzeemaneffect}

\begin{figure}
\begin{center}
\includegraphics[width=0.3\textwidth]{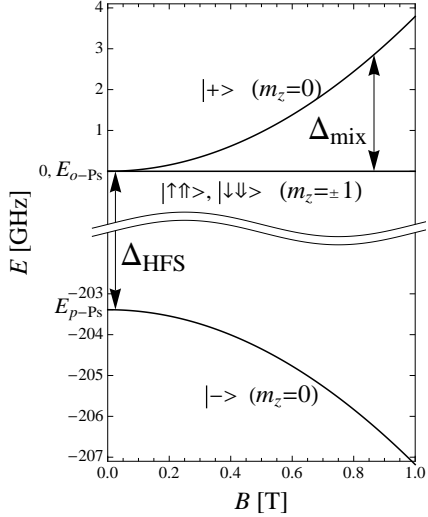}
\caption{\label{fig:pslevel}Zeeman energy levels of Ps in its ground state. The arrows $\uparrow, \downarrow$ means 
the spin up and down of electron, and the arrows $\Uparrow, \Downarrow$ means the spin up and down of positron.}
\end{center}
\end{figure}

The energy levels of the ground state of Ps are shown as a function of static magnetic field 
in Figure \ref{fig:pslevel}. Due to technical difficulties in directly stimulating 
$\Delta _{\mathrm{HFS}}$, we make an indirect measurement by stimulating the 
transition $\Delta _{\mathrm{mix}}$. This is the same approach as previous experiments. 
The relationship between $\Delta _{\mathrm{HFS}}$ and $\Delta _{\mathrm{mix}}$ is 
approximately given by the Breit-Rabi equation 
\begin{equation}
\Delta _{\mathrm{mix}} \simeq \frac{1}{2} \Delta _{\mathrm{HFS}} \left( \sqrt{1+4x^2} - 1 \right) \, ,
\label{eq:Zeeman}
\end{equation}
in which $x=g^{\prime}\mu _B B  / h\Delta _{\mathrm{HFS}}$, $g^{\prime} = g\left( 1-\frac{5}{24} \alpha ^2 \right)$ 
is the $g$ factor for a positron (electron) in Ps~\cite{G-FACTOR}, 
$\mu _B$ is the Bohr magneton, $B$ is the static magnetic field, and $h$ is the Plank constant.

In a static magnetic field, the $|S, m_z \rangle = |0,0 \rangle $, 
where $S$ is the total spin of Ps and $m_z$ is 
the magnetic quantum number of Ps along with z-axis (direction of static magnetic field), 
state mixes with the $|1,0 \rangle$ state 
hence the $|+\rangle$ state annihilates into 2 $\gamma$-rays with a lifetime of about 
8\,ns (with our experimental conditions).
The $|1, \pm 1 \rangle$ states annihilate into 3 $\gamma$-rays with a lifetime of 
about 140\,ns. When a microwave field with a frequency of $\Delta _{\mathrm{mix}}$ is 
applied, transitions between the $|+\rangle$ state and the $|1, \pm 1 \rangle$ states are 
induced so that the 2 $\gamma$-ray annihilation rate increases and the 3 $\gamma$-ray 
annihilation rate decreases. 
This change of annihilation rates is our experimental signal.

Our experimental resonance line shape is obtained using density matrix. 
We use the basis for four spin eigenstates of Ps as $\left( \psi _0, \psi _1, 
\psi _2, \psi _3 \right) \equiv \left( | 0, 0 \rangle , 
| 1, 0\rangle , | 1, 1\rangle , | 1, -1 \rangle \right) $. We apply a magnetic field 
\begin{equation}
 \mbox{\boldmath $B$} (t) = B {\bf e}_{\mathrm{z}} + B_0 {\bf e}_{\mathrm{x}} \cos \left( \omega t \right)\,,
\end{equation}
where ${\bf e}_{\mathrm{z}}$, ${\bf e}_{\mathrm{x}}$ are the unit vectors for z, x direction respectively, 
$B_0$ is magnetic field strength of microwaves, and $\omega$ is the frequency of microwaves. 
Then the Hamiltonian $\mbox{\boldmath $H$}$ becomes
\begin{equation}
\mbox{\boldmath $H$} = \left(
\begin{array}{cccc}
-\frac{1}{2} - \frac{\mathrm{i}}{2} \gamma _{\mathrm{s}} & x & -y & y \\
x & \frac{1}{2}-\frac{\mathrm{i}}{2} \gamma _{\mathrm{t}} & 0 & 0 \\
-y & 0 & \frac{1}{2}-\frac{\mathrm{i}}{2}\gamma _{\mathrm{t}} & 0 \\
y & 0 & 0 & \frac{1}{2}-\frac{\mathrm{i}}{2}\gamma _{\mathrm{t}} 
\end{array}
\right) h \Delta _{\mathrm{HFS}} \, ,
\end{equation}
where $y = C_y \frac{ g^{\prime } \mu _B B_0 }{h \Delta _{\mathrm{HFS}}} \cos \left( \omega t \right)$, 
$C_y$ is a constant, 
$\gamma _{\mathrm{s}} = \frac{ \Gamma _{\textrm{p-Ps}}}{2\pi \Delta _{\mathrm{HFS}}}$, 
$\gamma _{\mathrm{t}} = \frac{ \Gamma _{\textrm{o-Ps}}}{2\pi \Delta _{\mathrm{HFS}}}$, 
$\Gamma _{\textrm{p-Ps}}$ is the decay rate of p-Ps, and $\Gamma _{\textrm{o-Ps}}$ is that of o-Ps. 
The most recent and precise experimental values are 
$\Gamma _{\textrm{o-Ps}} = 7.040\,1(7)\,\mu\mathrm{s}^{-1}$~\cite{KATAOKA} and 
$\Gamma _{\textrm{p-Ps}} = 7.990\,9(17)\,\mathrm{ns}^{-1}$~\cite{GAMMA-0}. 

From the time-dependent Schr\"{o}dinger equation, the $4\times 4$ density matrix $\mbox{\boldmath $\rho$}(t)$ is given by 
\begin{equation}
\label{eq:rho}
\mathrm{i} \hbar \dot{\mbox{\boldmath $\rho$} } = \mbox{\boldmath $H$} \mbox{\boldmath $\rho$} -
 \mbox{\boldmath $\rho$} \mbox{\boldmath $H$} ^{\dagger } \, ,
\end{equation}
where the i,j-element of $\mbox{\boldmath $\rho$}(t)$ is defined as 
$\rho _{\mathrm{ij}} (t) \equiv \langle \psi _{\mathrm{i}} | 
\psi (t) \rangle \langle \psi (t) | \psi _{\mathrm{j}} \rangle $. 
If we take the initial state to be unpolarized, $\mbox{\boldmath $\rho$} (0) = \mathrm{diag} \left(\frac{1}{4}, 
\frac{1}{4}, \frac{1}{4}, \frac{1}{4} \right)$.

The 2$\gamma$-ray annihilation probability $S_{2\gamma}$ and 3$\gamma$-ray annihilation probability $S_{3\gamma}$, 
between $t=t_0$ and $t=t_1$, are obtained by
\begin{equation}
\label{eq:s2gamma}
S_{2\gamma} = \Gamma _{\textrm{p-Ps} } \int _{t_0} ^{t_1} \rho _{00}(t) dt\,,
\end{equation}
\begin{equation}
\label{eq:s3gamma}
S_{3\gamma } = \Gamma _{\textrm{o-Ps} } \int _{t_0} ^{t_1} 
\left( \rho _{11}(t) + \rho _{22}(t) + \rho _{33}(t) \right) dt\, .
\end{equation}

\subsection{Ps Thermalization Effect}
\label{sec:psthermalizationeffect}
Forming Ps needs material to provide electrons, but 
material around Ps makes electric field and changes $\Delta _{\mathrm{HFS}}$. 
This effect is called the Stark effect. 
The material effect must be properly considered to evaluate $\Delta _{\mathrm{HFS}}$ in vacuum. 
In the previous experiments, 
the material effect on $\Delta _{\mathrm{HFS}}$ was considered to be proportional to the material (gas) density. 
$\Delta _{\mathrm{HFS}}$ was measured at various gas density, and they were extrapolated linearly to zero 
density~\cite{HUGHES-V, MILLS-II}. 
But this extrapolation method can make large systematic error enough to account for the 
discrepancy between $\Delta _{\mathrm{HFS}} ^{\mathrm{exp}}$ 
and $\Delta_{\mathrm{HFS}} ^{\mathrm{th}}$. 

Formed Ps has the initial energy of O(eV). 
Positronium loses its energy when it collides with material, and finally its energy becomes to 
room temperature ($\sim1/30\,\mathrm{eV}$). 
This process is called thermalization. 
The material effect between $t=t_0$ and $t=t_1$ is proportional to 
\begin{equation}
\label{eq:materialeffect}
 \int _{t_0} ^{t_1} f(t) \left( \rho _{00} (t) (B_0 \neq 0) - \rho _{00} (t) (B_0 = 0) \right) dt\,, 
\end{equation}
where $f(t) \sim n \sigma v(t) $ is the collision rate of Ps with material, 
$n$ is the number density of material, $\sigma $ is the typical cross section of collision, 
and $v(t)$ is the mean velocity of Ps. In the previous experiments, 
timing information was not measured so that $t_0 = 0$ and $t_1 = \infty$. 
The effect is proportional to the material density if the thermalization occurs much faster than the o-Ps lifetime, 
but the thermalization time scale becomes large especially at low material density, which 
makes nonlinear effect on $\Delta _{\mathrm{HFS}}$. 

The nonlinear effect can be estimated using the thermalization model~\cite{NETSUKA-3}, 
\begin{equation}
\frac{ dE_{\mathrm{av}}(t)}{dt} = 
- \sqrt{2 m_{\mathrm{Ps}}E_{\mathrm{av}}(t)}\left(E_{\mathrm{av}}(t) - \frac{3}{2}k T \right)
 \frac{8}{3}\sqrt{\frac{2}{3\pi}} \frac{2\sigma _{\mathrm{m}} n}{M} \,,
\end{equation}
where $E_{\mathrm{av}}(t)$ is the average Ps energy, $m_{\mathrm{Ps}}$ is the Ps mass, 
$M$ is the mass of the gas molecule, $T$ is the temperature of the gas, $k$ is the Boltzmann constant, and 
$ \frac{8}{3}\sqrt{\frac{2}{3\pi}} \frac{2\sigma _{\mathrm{m}} n}{M}$ means the collision effect. 
The solution of this equation is~\cite{NETSUKA-2} 
\begin{equation}
\label{eq:thermalization}
E_{\mathrm{av}}(t) = \left( \frac{1+A e^{-bt}}{1-Ae^{-bt}} \right) ^2 \frac{3}{2}k T\,,
\end{equation}
where $b = \frac{8}{3}\sqrt{\frac{2}{3\pi}} \frac{2\sigma _{\mathrm{m}} n }{M} \sqrt{ 3m_{\mathrm{Ps}} k T}$, 
$ A = \frac{ \sqrt{E_0} - \sqrt{ \frac{3}{2} k T }}{ \sqrt{E_0} + \sqrt{ \frac{3}{2} k T } }$, 
and $E_0 \equiv E_{\mathrm{av}}(0)$ is the initial energy of Ps. 

\begin{figure}
\begin{center}
\includegraphics[width=0.4\textwidth]{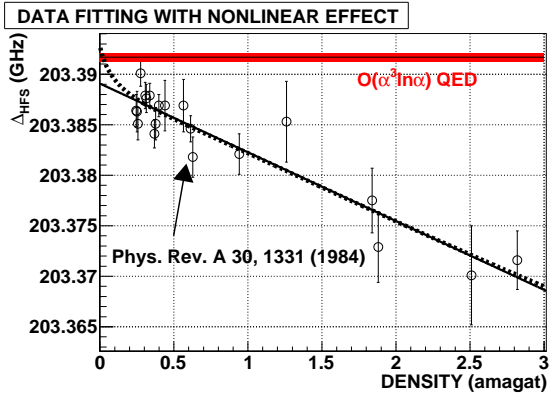}
\caption{\label{fig:nonlinear}(Color online) Thermalization effect on $\Delta _{\mathrm{HFS}}$. 
The circles and error bars are the data of Ref.~\cite{HUGHES-V}, the solid line is the linear fit, 
the dashed line is the best fit including Ps thermalization effect, and the red band is the 
$\mathrm{O} (\alpha ^3 \ln \alpha )$ QED prediction in vacuum.}
\end{center}
\end{figure}

Figure \ref{fig:nonlinear} shows the best fit result of the previous experimental data from Ref.~\cite{HUGHES-V} 
considering nonlinear material effect using Equations (\ref{eq:materialeffect}) and (\ref{eq:thermalization}) 
with parameters $E_0 = 2.07\,\mathrm{eV}$ and $\sigma _{\mathrm{m}} = 13.0$\,\AA$^2$~\cite{NETSUKA-5},  
compared with linear fitting. The nonlinear effect can be clearly seen at low density, and it can 
explain the discrepancy between $\Delta _{\mathrm{HFS}} ^{\mathrm{exp}}$ 
and $\Delta_{\mathrm{HFS}} ^{\mathrm{th}}$ (the best fit value including nonlinear effect is 
$\Delta _{\mathrm{HFS}} = 203.392\,80(95)\,\mathrm{GHz}$). 
But $\sigma _{\mathrm{m}} = 37 \pm 10$\,\AA$^2$ from Ref.~\cite{NETSUKA-3}, which is not consistent 
with the value of Ref.~\cite{NETSUKA-5}, affects only about 5\,ppm and cannot explain the discrepancy. 
Therefore, new independent measurement of Ps thermalization is needed. 

\subsection{Our new methods}
\label{sec:ournewmethods}
Our new methods will significantly reduce the systematic errors present in previous experiments; 
the Ps thermalization effect and the non-uniformity of the magnetic field. 
The main improvements in our experiment are the large bore superconducting magnet, 
$\beta$-tagging system, and high performance $\gamma$-ray detectors. 
Details are discussed in the following sections. 

\section{Prototype Run}
\label{sec:prototyperun}
The prototype run of the measurement with new methods has been performed.  

\subsection{Experimental Setup}
\label{sec:experimentalsetup}
\begin{figure}
\begin{center}
\includegraphics[width=0.48\textwidth]{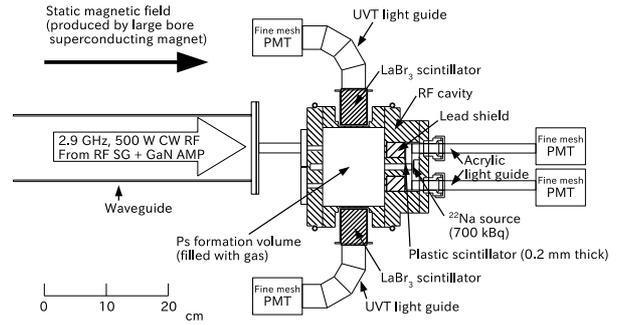}
\caption{\label{fig:schematic}Schematic diagram of the experimental setup of the prototype run (top view in magnet).}
\end{center}
\end{figure}

A schematic diagram of the experimental setup of the prototype run 
is shown in Figure~\ref{fig:schematic}.

\subsubsection{Large bore superconducting magnet}
\label{sec:largeboresuperconductingmagnet}
A large bore superconducting magnet is 
used to produce the magnetic field $B \sim 0.866\,\mathrm{T}$ which induces the Zeeman splitting. 
The bore diameter of the magnet is 800\,mm, and its length is 2\,m. 
The large bore diameter means that there is good uniformity in the magnetic field in the 
region where Ps is formed. 
Furthermore, the magnet is operated in persistent current mode, making the stability of the magnetic field 
better than $\pm$1\,ppm.

The magnetic field distribution was measured using a proton NMR magnetometer (ECHO Electronics EFM-150HM-AX) 
with a clock synthesizer (NF Corporation CK1615 PA-001-0312). Calibration uncertainty of the system is 
less than 1\,ppm. Non-uniformity of the magnetic field in a whole volume of RF cavity is 
23.1\,ppm (RMS) without any compensation coils. The non-uniformity decreases to 10.4\,ppm when weighted by 
RF power and positron stop position distribution. 

\subsubsection{$\beta$-tagging system and timing information}
\label{sec:betataggingsystemandtiminginformation}
The positron source is 19\,$\mu$Ci (700\,kBq) of $ \mathrm{^{22}Na}$ (Eckert \& Ziegler POSN-22). 
A plastic scintillator (NE102A) 10\,mm in diameter and 0.2\,mm thick is used to tag positrons emitted from the $\mathrm{^{22}Na}$.
The scintillation light is detected by fine mesh photomultiplier tubes (PMT: HAMAMATSU H6614-70MOD) 
and provides a start signal which corresponds to the time of Ps formation. 
The timing resolution is 1.0\,ns (1$\sigma$).
The positron then enters the microwave cavity, forming Ps in the gas contained therein. 

Ps decays into photons that are detected with LaBr$_3$\,(Ce) scintillators (Saint-Gobain BrilLanCe$^{\mathrm{TM}}$ 380). 
Accumulating measurements of the times of positron emission and $\gamma$-detection results in 
decay curves of Ps as shown in Figure \ref{fig:timing_spectra}.
The timing information is used to improve the accuracy of the measurement of $\Delta _{\mathrm{HFS}}$ as follows: 
\begin{enumerate}
\item Imposing a time cut means that we can select well thermalized Ps, reducing the unthermalized o-Ps contribution. 
It should also be possible to precisely measure the contributions of 
unthermalized o-Ps, and of material effects (we plan to make such measurements in future runs). 
\item A time cut also allows us to avoid the prompt peak (contributions of 
simple annihilation and of fast p-Ps decay), which greatly increases the S/N of the measurement (by about a factor of 20). 
\end{enumerate}

\subsubsection{High performance $\gamma$-ray detectors}
\label{sec:highperformancegammaraydetectors}
Six $\gamma$-ray detectors are located around the microwave cavity to detect 
the 511\,keV annihilation $\gamma$-rays. 
LaBr$_3$ scintillators, 1.5 inches in diameter and 2 inches long are used. 
The scintillation light is detected by fine mesh PMT through the UVT light guide. 
Without light guide, LaBr$_3$ scintillators have 
good energy resolution (4\% FWHM at 511\,keV) and timing resolution (0.2\,ns FWHM at 511\,keV), 
and have a short decay constant (16\,ns). 
The good energy resolution and the high counting rate of LaBr$_3$ results in very good overall performance 
for measuring Zeeman transitions. 
In particular the good energy resolution allows us to efficiently separate 2$\gamma$ events from 
3$\gamma$ events, negating the need to use a back-to-back geometry to select 2$\gamma$ events, thus 
greatly increasing the acceptance of our setup. 

This $\gamma$-ray detector system greatly reduces the statistical error in the measurement. 

\subsubsection{RF system}
\label{sec:rfsystem}
Microwaves are produced by 
a local oscillator signal generator (ROHDE \& SCHWARZ SMV 03) and amplified to 500\,W with a 
GaN amplifier (R\&K A2856BW200-5057-R).
 
The microwave cavity is made with oxygen-free copper; 
the inside of the cavity is a cylinder 128\,mm in diameter and 
100\,mm long. The side wall of the cavity is only 2\,mm thick in order to 
allow the $\gamma$-rays to efficiently escape. 
The cavity is operated in the TM$_{110}$ mode. The resonant frequency is 2.856\,6\,GHz and 
$Q_L = 14,700 \pm 50$. The cavity is filled with gas 
(90\% N$_2$ and 10\% iso-C$_4$H$_{10}$) with a gas-handling system. Iso-C$_4$H$_{10}$ is used as the quenching gas 
to remove background 2 $\gamma$-ray annihilation.

\subsubsection{Monte Calro simulation}
\label{sec:montecalrosimulation}
The Monte Calro simulation to use in analysis is performed using Geant4~\cite{GEANT4}. 
The low energy physics package PENELOPE~\cite{PENELOPE} is used and the geometry of the experimental setup is 
carefully input. 
The simulation is produced at each magnetic field strength and gas density. 

\subsubsection{Data acquisition}
\label{sec:dataacquisition}
The prototype run was performed from 2 July 2009 to 24 September 2009 
using the large bore magnet with no compensation (compensation magnets to reduce 
the uniformity to O(ppm) are planned but are not yet installed). 
In the overall period, the trigger rate was about 3.6\,kHz and the data acquisition rate was about 
650\,Hz. The data acquisition was performed using NIM and CAMAC system. 

The trigger signal is the coincidence signal of $\beta$-tagging system and $\gamma$-ray detectors. 
Timing information of all PMT are obtained with a 2\,GHz direct clock counting Time-to-Digital Converter 
(TDC: GNC-060)~\cite{KATAOKA, KATAOKA-D}. 
A charge ADC (CAEN C1205) is used to measure the energy information of the LaBr$_3$ crystals while 
another charge ADC is used to measure the energy information of the plastic scintillators. 
A crate controller (TOYO CC/NET) is used and obtained data are stored in a HDD of a Linux PC via Ethernet. 
The Zeeman transition has been measured at various magnetic field strengths with a fixed RF frequency and power.
The transition resonance lines are obtained at two gas densities (1.350\,1\,amagat and 0.891\,6\,amagat).

\subsection{Analysis}
\label{sec:analysis}

\subsubsection{Data analysis}
\label{sec:dataanalysis}

\begin{figure}
\begin{center}
\includegraphics[width=0.4\textwidth]{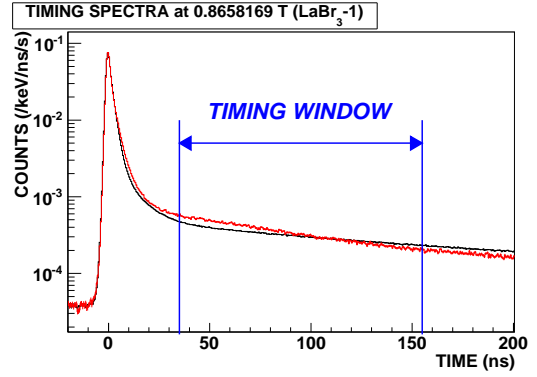}
\caption{\label{fig:timing_spectra}(Color online) Decay curves of Ps. The red line is on-resonance (0.865\,816\,9\,T) RF\,ON, and the black line is RF\,OFF. 
The timing window of 35--155\,ns is also indicated. The decay rate of Ps increases with RF because of the Zeeman transition.}
\end{center}
\end{figure}

\begin{figure}
\begin{center}
\subfigure[Typical fitting of energy spectrum (RF-OFF). 
The black points are the data, the red line is $S_{2\gamma} ^{\mathrm{MC}}$, the blue line is $S_{3\gamma } ^{\mathrm{MC}}$, 
and the pink line is sum of the Monte Calro spectra.]{
	\includegraphics[width=0.4\textwidth]{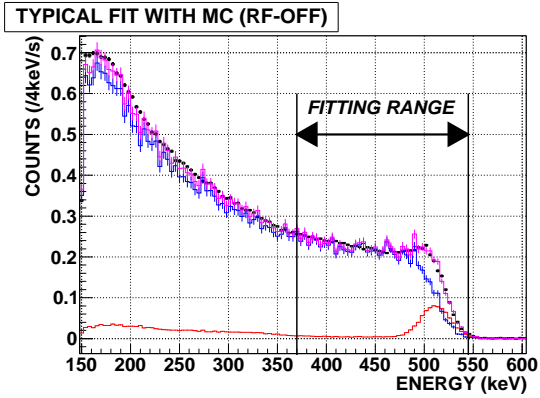}
	\label{fig:energy_spectraa}}
\subfigure[Typical fitting of energy spectrum (RF-ON). 
The black points are the data, the red line is $S_{2\gamma ,\mathrm{in}} ^{\mathrm{MC}}$, 
the blue line is $S_{3\gamma ,\mathrm{in}} ^{\mathrm{MC}}$, 
the green line is $S_{2\gamma ,\mathrm{out}} ^{\mathrm{MC}}$, 
the aqua line is $S_{3\gamma ,\mathrm{out}} ^{\mathrm{MC}}$, 
and the pink line is sum of the Monte Calro spectra. $A_{\mathrm{OFF}}$ and $\beta $ are 
fixed with the fitted values of RF-OFF, so that the free parameters are $A_{\mathrm{ON}}$ and $\Gamma $.]{
	\includegraphics[width=0.4\textwidth]{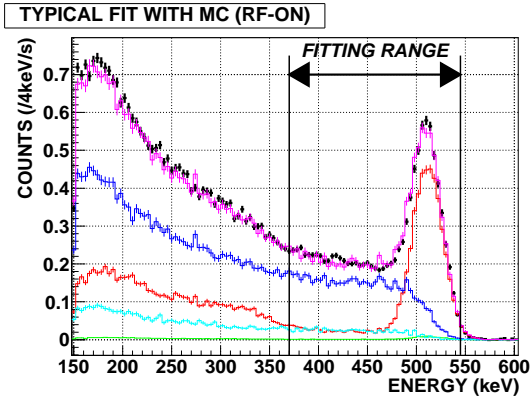}
	\label{fig:energy_spectrab}}
	\caption{(Color online) Typical energy spectra and its fitting with Monte Calro simulation.
The data at 1.350\,1\,amagat, 0.865\,816\,9\,T are shown and the spectra are normalized by the live time.}
	\label{fig:energy_spectra}
\end{center}
\end{figure}

Figure \ref{fig:timing_spectra} shows examples of measured timing spectra. The prompt peak coming from annihilation and 
$|-\rangle$ decay is followed by the decay curve of $|+\rangle$ and $m_z=\pm 1$ states, then the constant accidental spectrum. 
A timing window of 35--155 ns is applied to select transition events. 
Figure \ref{fig:energy_spectra} shows energy spectra, which are obtained by 
subtracting the accidental contribution using the timing window $t = $700--850\,ns.

The energy spectra are fitted using Monte Calro simulation. 
The energy spectrum of RF-OFF $S_{\mathrm{OFF}}$ can be expressed as 
\begin{equation}
S_{\mathrm{OFF}} = A_{\mathrm{OFF}} \left( S_{\mathrm{3\gamma}} ^{\mathrm{MC}} + 
\beta  S_{\mathrm{2\gamma}} ^{\mathrm{MC}} \right) \, ,
\end{equation}
where $A_{\mathrm{OFF}}$ is a normalizing constant, $\beta $ is a $2\gamma / 3\gamma $ decay ratio of RF-OFF, 
$S_{\mathrm{2\gamma}} ^{\mathrm{MC}}$ is a Monte Calro simulated energy spectrum of $2\gamma $ decay, 
and $S_{\mathrm{3\gamma }} ^{\mathrm{MC}} $ is that of $3\gamma $ decay.
On the other hand, the energy specrum of RF-ON $S_{\mathrm{ON}}$ can be expressed as 
\begin{equation}
S_{\mathrm{ON}} = A_{\mathrm{ON}} \left( S_{\mathrm{3\gamma}, \mathrm{in}} ^{\mathrm{MC}} + 
\Gamma  S_{\mathrm{2\gamma}, \mathrm{in}} ^{\mathrm{MC}} \right) +
A_{\mathrm{OFF}} \left( S_{\mathrm{3\gamma}, \mathrm{out}} ^{\mathrm{MC}} + 
\beta  S_{\mathrm{2\gamma}, \mathrm{out}} ^{\mathrm{MC}} \right) \, ,
\end{equation}
where $A_{\mathrm{ON}}$ is a normalizing constant and $\Gamma $ is a $2\gamma / 3\gamma $ decay ratio of RF-ON. 
The subscript ``in" and ``out" means the spectrum of $\gamma$-rays from Ps decays in the volume where RF power is applied 
or not applied, respectively. 

Typical fitted results are also shown in Figure \ref{fig:energy_spectra}. Figure \ref{fig:energy_spectraa} is RF-OFF, 
and \ref{fig:energy_spectrab} is RF-ON. 
The energy spectrum is measured at different magnetic field strengths. 
The fitting range is 370--545\,keV. 
All the spectra are fitted successfully using MINUIT~\cite{MINUIT}. 

\begin{figure}
\begin{center}
\includegraphics[width=0.4\textwidth]{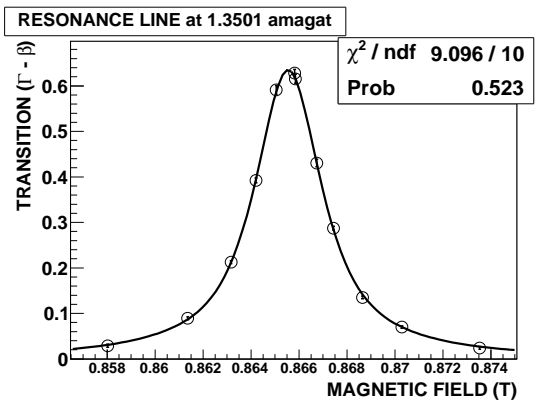}
\caption{\label{fig:resonance_line}Resonance line at 1.350\,1 amagat gas density. The circles and error bars are the data, and 
the solid line is the best fit result. 
The error bars include errors from statistics of data, statistics of Monte Calro simulation, uncertainty of RF power, and uncertainty of 
$Q_L$ value of the RF cavity.}
\end{center}
\end{figure}

We take $\Gamma - \beta $ as an amount of Zeeman transition. 
An example of resonance line obtained is shown in Figure \ref{fig:resonance_line}. 
Resonance lines can be fitted by 
$S_{2\gamma }/S_{3\gamma } (\textrm{RF-ON}) - S_{2\gamma }/S_{3\gamma } (\textrm{RF-OFF})$ .
The theoretical function is calculated from Equations (\ref{eq:s2gamma}) and (\ref{eq:s3gamma}) numerically 
using RKF45 formula, with contributions from pick-off and slow positron. 
The typical $B_0$ is 14.2\,G. The RF-OFF function is obtained by substitute 0 for $B_0$. 
Then the free parameters of fitting are $\Delta _{\mathrm{HFS}}$ and $C_y$. 
The fitting results are summarized in Table~\ref{tab:fittingresult}. 
We use $-33\,\mathrm{ppm/amagat}$~\cite{HUGHES-V} as the material effect and 
obtain $\Delta _{\mathrm{HFS}}$ in vacuum. 

\begin{center}
\begin{table}
\caption{\label{tab:fittingresult}Fitting result of the resonance lines. These uncertainties include 
errors from statistics of data, statistics of Monte Calro simulation, uncertainty of RF power, and uncertainty of 
$Q_L$ value of the RF cavity.}
\begin{center}
\begin{tabular}{crrr}
\hline \hline
Gas density & $\Delta _{\mathrm{HFS}}$ & Relative error & $\chi ^2 / \mathrm{ndf}$ \\
(amagat) & (GHz) & (ppm) & \\
\hline
1.350\,1(71) & 203.368\,3(55) & 27 & 0.910 \\
0.891\,6(23) & 203.379\,3(70) & 34 & 0.483 \\
\hline \hline
\end{tabular}
\end{center}
\end{table}
\end{center}

\subsubsection{Systematic Errors}
\label{sec:systematicerrors}
Systematic errors of the prototype run are summarized in Table~\ref{tab:systematicerror}.
\begin{enumerate}
\item {\it Magnetic Field}. The largest uncertainty in the prototype run is non-uniformity of the 
magnetic field. The weighted non-uniformity is 10.4\,ppm. 
The offset and reproducibility of the magnetic field is measured to be 2\,ppm. 
The calibration uncertainty of NMR magnetometer is 1\,ppm. 
These uncertainties are doubled because 
$\Delta _{\mathrm{HFS}}$ is approximately proportional to square of the magnetic field strength. 
\item {\it Monte Calro Simulation}. The magnetic field dependence of $\beta$ is not 
exactly reproduced by the energy spectra fitting with Monte Calro simulation. 
From the difference between theoretical dependence and fitted result, this effect is 
estimated to be 18\,ppm. The statistical uncertainty of Monte Calro simulation is 17\,ppm. 
\item {\it RF System}. The uncertainty of $Q_L$ value of RF cavity comes from 
reproducibility of the RF environment ($\sim 5\%$) and 
uncertainty of the $Q_L$ measurement method (0.6\%). Its effect on $\Delta _{\mathrm{HFS}}$ is 
estimated to be 6\,ppm. 
RF power uncertainty comes from reproducibility of the RF environment ($\sim 2\%$), 
uncertainty of power meters (0.05\%), and uncertainty of the power meter calibration of 
temperature dependence (0.03\%/K). It contributes to the error on $\Delta _{\mathrm{HFS}}$ by 
5\,ppm. 
The uncertainty of RF frequency is 5\,ppm. It directly affects $\Delta _{\mathrm{HFS}}$ because 
$\Delta _{\mathrm{HFS}}$ is approximately proportional to inverse of RF frequency. 
\item {\it Material Effect}. Thermalization of Ps can affect $\Delta _{\mathrm{HFS}}$ by up to 20\,ppm, but 
it has been not yet measured. Gas density dependence has not been measured in the prototype run, so 
we have used the value from previous experiment. The uncertainty of the density dependence is 
4\,ppm/amagat~\cite{HUGHES-V}, and the uncertainty from iso-C$_{4}$H$_{10}$ is estimated to be less than 7.7\,ppm/amagat. 
These result in 7\,ppm uncertainty of $\Delta_{\mathrm{HFS}}$. 
\end{enumerate}
Other uncertainties are considered to be negligible. The systematic errors discussed above are summed in quadrature. 

\begin{center}
\begin{table}
\caption{\label{tab:systematicerror}Summary of Systematic errors of the prototype run.}
\begin{center}
\begin{tabular}{lr}
\hline \hline
Source & Errors in $\Delta _{\mathrm{HFS}}$ (ppm) \\
\hline
{\it Magnetic Field:} & \\
~~Non-uniformity & 21 \\
~~Offset and reproducibility & 4 \\
~~NMR measurement & 2 \\
{\it Monte Calro Simulation:} & \\
~~Magnetic field dependence & 18 \\
~~Statistics & 17 \\
{\it RF System:} & \\
~~$Q_L$ value of RF cavity & 6 \\
~~RF power & 5 \\
~~RF frequency & 5 \\
{\it Material Effect:} & \\
~~Thermalization of Ps & $<20$ \\
~~Gas density dependence & 7 \\
\hline
Quadrature sum & 40 \\
\hline \hline
\end{tabular}
\end{center}
\end{table}
\end{center}

\subsection{Result}
\label{sec:result}

The value of $\Delta_{\mathrm{HFS}}$ obtained from the prototype run is 
\begin{eqnarray}
\Delta _{\mathrm{HFS}} &=& 203.380\,4 \pm 0.002\,2 (\mathrm{stat.}, 11\,\mathrm{ppm}) \nonumber \\
& & \pm 0.008\,1 (\mathrm{sys.},40\,\mathrm{ppm})
\, \mathrm{GHz}\, ,
\end{eqnarray}
which is consistent with both of the previous experimental values and with the theoretical value. 
The uncertainty is an order of magnitude larger than the goal precision of O(ppm). 

\subsection{Next steps}
The following improvements are planned for future measurements:
\begin{enumerate}
\item Compensation magnets will be installed and O(ppm) magnetic field uniformity is 
expected to be achieved. 
\item The Monte Calro simulation will be studied to reproduce the magnetic field dependence of 
energy spectra of Ps decays. Statistics of simulation will be reduced to O(ppm). 
\item The errors from RF system will be reduced to O(ppm) by carefully controlling the 
environment (especially the temperature) of the experiment. 
\item Measurements at various pressures of gas will be performed to estimate 
the material effect (the Stark effect). The accumulation of these measurements will 
result in an O(ppm) statistical error within a few years.
\item The timing information allows for a measurement of Ps thermalization as a function of 
time~\cite{KATAOKA, KATAOKA-D, ASAI}. We can thus precisely measure the material effect including the 
thermalization effect. 
\end{enumerate}

\section{Conclusion}
\label{sec:conclusion}
A new experiment to measure the Ps-HFS which reduces possible common uncertainties in previous experiments 
has been constructed and the prototype run has been finished. A value of 
$\Delta _{\mathrm{HFS}} = 203.380\,4 \pm 0.002\,2 (\mathrm{stat.}) \pm 0.008\,1 (\mathrm{sys.}) \, \mathrm{GHz} \, (41\,\mathrm{ppm})$ has been obtained, which is 
consistent with both of the previous experimental values and with the theoretical calculation. Development of compensation magnets 
is underway with a view to obtaining O(ppm) magnetic field homogeneity for the final run. 
The final run will start soon. 
A new result with an accuracy of O(ppm) will be 
obtained within a few years which will be an independent check of the discrepancy between the present experimental values and the QED prediction.



%% References


\bibliographystyle{elsarticle-num}
%% \bibliography{<your-bib-database>}
\bibliography{ishida}






% \end{thebibliography}

\end{document}